\documentclass[twocolumn,prl]{revtex4}
\usepackage{graphicx}% Include figure files
\usepackage{dcolumn}% Align table columns on decimal point
\usepackage{bm}% bold math
\usepackage{multirow}
\usepackage[dvipsnames]{xcolor}
\usepackage{color}
\usepackage[normalem]{ulem}
\usepackage{amsmath}
\usepackage{gensymb}

\usepackage[draft]{changes}

\usepackage[colorlinks=true]{hyperref}
\begin{document}

\title{Control of light emission of quantum emitters coupled to silicon nanoantenna using cylindrical vector beams}

%%=============================================================%%

\author{Martin Montagnac$^{1}$}
\author{Yoann Br\^ul\'e$^{2}$}
\author{Aur\'elien Cuche$^{1}$}
\author{Jean-Marie Poumirol$^{1}$}
\author{S\'ebastien J. Weber$^{1}$}
\author{Jonas M\"uller$^{3}$}
\author{Guilhem Larrieu$^{3}$}
\author{Vincent Larrey$^{4}$}
\author{Franck Fournel$^{4}$}
\author{Olivier Boisron$^{5}$}
\author{Bruno Masenelli$^{6}$}
\author{G\'erard Colas des Francs$^{2}$}
\author{Gonzague Agez$^{1}$}\email{gonzague.agez@cemes.fr}
\author{Vincent Paillard$^{1}$}\email{vincent.paillard@cemes.fr}

%%=============================================================%%

\affiliation{\small$^1$CEMES-CNRS, Universit\'e de Toulouse, Toulouse, France}
\affiliation{\small$^2$ICB, Université de Bourgogne-Franche Comté, CNRS, Dijon, France}
\affiliation{\small$^3$LAAS-CNRS, Universit\'e de Toulouse, 31000 Toulouse, France}
\affiliation{\small$^4$CEA-LETI, Universit\'e Grenoble-Alpes, Grenoble, France}
\affiliation{\small$^5$Université de Lyon, Université Lyon 1, CNRS UMR 5510, ILM, Villeurbanne , France}
\affiliation{\small$^6$Université de Lyon, INSA Lyon, CNRS, Ecole Centrale de Lyon, Université Lyon 1, CPE, UMR 5270, INL, Villeurbanne , France}

\begin{abstract}
Light emission of europium (Eu$^{3+}$) ions placed in the vicinity of optically resonant nanoantennas is usually controlled by tailoring the local density of photon states (LDOS). We show that the  polarization and shape of the excitation beam can also be used to manipulate light emission, as azimuthally or radially polarized cylindrical vector beam offers to spatially shape the electric and magnetic fields, in addition to the effect of silicon nanorings (Si-NRs) used as nanoantennas. The photoluminescence mappings of the Eu$^{3+}$ transitions and the Si phonon mappings are strongly dependent of both the excitation beam and the Si-NR dimensions. The experimental results of Raman scattering and photoluminescence are confirmed by numerical simulations of the near-field intensity in the Si nanoantenna and in the Eu$^{3+}$-doped film, respectively. The branching ratios obtained from the experimental PL maps also reveal a redistribution of the electric and magnetic emission channels. Our results show that it is possible to  spatially control both electric and magnetic dipolar emission of Eu$^{3+}$ ions by switching the laser beam polarization, hence the near-field at the excitation wavelength, and the electric and magnetic LDOS at the emission wavelength. This paves the way for optimized geometries taking advantage of both excitation and emission processes.
\end{abstract}

\maketitle

%% ------------------------ Section ------------------------
\section*{Introduction}
Enhancing and controlling the light emission of quantum emitters coupled to optically resonant nanostructures is essential for new light sources at the nanoscale, and for enhanced spectroscopies and biosensing \cite{krasnok_spectroscopy_2018}. The photoluminescence (PL) intensity of a quantum emitter at a position $\mathbf{r}$ coupled to a nanoantenna is influenced by three parameters, as expressed in the following equation \cite{koenderink_single-photon_2017,bidault_dielectric_2019}:

\begin{equation} 
\label{PL_intensity}
\begin{split}
I(\mathbf{r},\omega_{exc},\omega_{em}) = 
P_{exc}(\mathbf{r},\omega_{exc})~\times~\phi_{em}(\mathbf{r},\omega_{em})\\
~\times~C_{coll}(\mathbf{r},\omega_{em})
\end{split}
\end{equation}

where $P_{exc}(\mathbf{r},\omega_{exc})$ is the excitation rate, proportional to the excitation pump, $\phi_{em}(\mathbf{r},\omega_{em})$ is the quantum yield related to the local density of photon states (LDOS) at the emission angular frequency $\omega_{em}$, and $C_{coll} (\mathbf{r},\omega_{em})$ is the collection efficiency, that depends on both the emission directivity and detection geometry.\\
\indent Many works have been devoted to LDOS engineering \cite{li_enhancing_2017, sanz-paz_enhancing_2018, mignuzzi_nanoscale_2019, brule_magnetic_2022} or to emission directivity \cite{curto_unidirectional_2010, poumirol_unveiling_2020, wiecha_design_2019,humbert_large-scale_2023}, but the influence of the pump has been rarely considered experimentally. In fact, the PL signal of a quantum emitter is either proportional to the LDOS or to the near-field intensity when the excited state is saturated or not, respectively \cite{girard_generalized_2005, wiecha_enhancement_2019, majorel_quantum_2020}. In this article, we investigate in the nonsaturated regime the influence of focused cylindrical vector beams on the PL of a rare earth ion-doped thin film deposited on high refractive index dielectric nanostructures.\\
\indent Rare-earth ion-based quantum emitters have been widely studied over the last decade because they are photostable and present narrow electronic transitions, corresponding to efficient electric (ED) or magnetic (MD) dipole transitions, either at the absorption \cite{kasperczyk_excitation_2015} or at the emission \cite{karaveli_spectral_2011, aigouy_mapping_2014}. They have thus been used to probe the electric and magnetic components of light by placing them in the vicinity of plasmonic or dielectric antennas, where the electric and magnetic LDOS (E-LDOS and M-LDOS) can be adjusted and spatially separated \cite{aigouy_mapping_2014, mivelle_strong_2015, rabouw_europium-doped_2016, wiecha_enhancement_2019}. To some extent, enhancing electric or magnetic Purcell effect could lead to applications in telecommunication using near-infrared emitters such as Er$^{3+}$ \cite{kalinic_all-dielectric_2020}, or efficient visible light sources and lamp phosphors using Eu$^{3+}$ or Tb$^{3+}$ \cite{baranov_modifying_2017}. The MD transitions, interacting with the magnetic components of light \cite{kasperczyk_excitation_2015, baranov_modifying_2017, wiecha_decay_2018}, may be even more interesting when they are coupled to high index dielectric nanoantennas, as in such nanostructures the magnetic field is known to be strongly enhanced at the magnetic dipole or quadrupole resonance wavelength \cite{rolly_promoting_2012, albella_low-loss_2013, bakker_magnetic_2015, matsumori_silicon_2022}. Experimentally, a larger enhancement of the MD emitting transition compared to the ED emitting transition has been reported \cite{wiecha_enhancement_2019, sugimoto_magnetic_2021}. Theoretically, a few recent works predict a very strong enhancement of the magnetic Purcell effect, larger than 10$^{3}$, in case of nanostructure design and M-LDOS optimization \cite{rocco_giant_2020, brule_magnetic_2022}.

Nevertheless, in addition to the nanoantenna design for tailoring the LDOS and the directional behavior as expressed in equation \ref{PL_intensity}, the excitation beam is the third parameter that can be used to spatially and spectrally shape the intensity of any optically driven light source.  Cylindrical vector beams (CVB) of radial or azimuthal polarization  can very helpful as the electric and magnetic fields are spatially separated \cite{dorn_sharper_2003, kasperczyk_excitation_2015}, and can be used to selectively excite electric and magnetic resonance modes in dielectric nanoantennas \cite{wozniak_selective_2015, montagnac_engineered_2022}, as well as electric and magnetic dipole transitions in rare-earth ions \cite{kasperczyk_excitation_2015}.\\
We thus show in the following that CVB can be used to spatially engineer the PL intensity of Eu$^{3+}$-doped thin films deposited on Si nanorings, due to different controlled near-field hot spots.
% by linearly polarized Gaussian beam. 

%% ------------------------ Figure ------------------------
\begin{figure}
\includegraphics[width=0.4\textwidth]{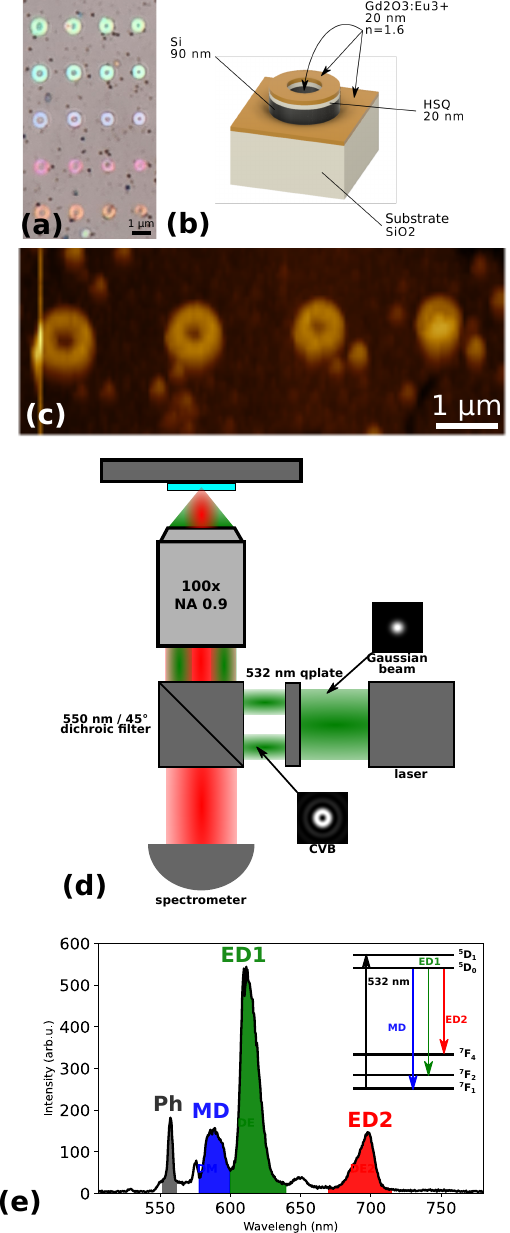}
\caption{\label{fig:fig1} \textbf{Experimental set-up and sample description.} (a) optical image of Si nanorings on SiO$_2$ substrate. (b) sketch showing the different layers. (c) AFM image of Eu$^{3+}$:Gd$_{2}$O$_{3}$ cluster-deposited film on Si nanorings. (d) sketch of the experimental set-up for PL imaging. (e) Typical spectrum taken on a Si-NR showing the Raman and PL signals; inset: sketch in energy of the transitions in Eu$^{3+}$ ions.}
\end{figure}
%% --------------------------------------------------------

%% ------------------------ Section ------------------------
\section*{Experimental set-up and numerical tools}

Si nanostructures are fabricated by electron beam lithography followed by reactive ion etching in a single crystal Si layer transferred on a fused silica (SiO$_2$) substrate, referred to as silicon on silica (SOS).  We chose nanorings (NRs) for their doughnut-like shape similar to the CVB symmetry. They are characterized by their outer radius $R$ and width $W$, their height  being fixed by the Si top layer of the SOS ($H$~=~90~nm).\\
In figure~\ref{fig:fig1}a, we show a bright field image of the Si nanoresonators covered by a 20 nm-thick film obtained by low energy cluster beam deposition (LECBD) of Eu$^{3+}$:Gd$_{2}$O$_{3}$ clusters \cite{wiecha_enhancement_2019}. The film homogeneity is very good as shown by atomic force microscope (AFM) image (see figure~\ref{fig:fig1}c). The distance between the Si-NR top surface and the Eu$^{3+}$:Gd$_{2}$O$_{3}$ deposit is about 20 nm, corresponding to the thickness of the hydrogen silsesquioxane (HSQ) resist, which was left on top of the Si-NRs during the fabrication process to protect the smallest structures from etching or lift-off (see figure~\ref{fig:fig1}b). The HSQ resist optical constants are the same as the silica ones. 

The photoluminescence (PL) mappings are obtained by raster scanning the sample under a laser beam tightly focused through a microscope objective of numerical aperture $NA=0.9$. 
%The excitation wavelength was either 528 nm or 532 nm, nearly resonant with the $^{7}$F$_{0}$ $\xrightarrow{}$ $^{5}$D$_{1}$ magnetic dipole (MD) transition or the $^{7}$F$_{1}$ $\xrightarrow{}$ $^{5}$D$_{1}$ electric dipole (ED) transition in Eu$^{3+}$, respectively.
The excitation wavelength of 532 nm is resonant with the $^{7}$F$_{1}$ $\rightarrow{}$ $^{5}$D$_{1}$ electric dipole (ED) transition in Eu$^{3+}$, and with a resonance mode of the Si-NRs (see supplemental information, Fig. S1) \cite{kasperczyk_excitation_2015}.
The linear polarization of the Gaussian excitation beam, referred to as lin-CVB, can be modified by a half-wave plate. The doughnut beam shape associated to azimuthal or radial polarization (referred to as azi-CVB and rad-CVB, respectively) is obtained by adding a Q-plate after the half-wave plate and before the microscope (figure~\ref{fig:fig1}d). We verified that the chosen polarization is conserved after reflection on the dichroic mirror.\\
\indent In all cases, the power density on the sample was kept low enough to stay below the saturation regime of the excited state (see supplementary information), thus avoiding the PL being governed by the LDOS only. In the low power regime, as previously mentioned, we expect the PL to be dependent mainly on the near-field at the excitation wavelength  \cite{girard_generalized_2005, wiecha_enhancement_2019, majorel_quantum_2020}.\\
A typical PL spectrum is displayed in figure~\ref{fig:fig1}. We will focus on three main features corresponding to $^{5}$D$_{0}$ $\xrightarrow{}$ $^{7}$F$_{1}$ centered around 590 nm (MD transition), $^{5}$D$_{0}$ $\xrightarrow{}$ $^{7}$F$_{2}$ around 610 nm (ED1 transition), and $^{5}$D$_{0}$ $\xrightarrow{}$ $^{7}$F$_{4}$ around 690 nm (ED2 transition) \cite{aigouy_mapping_2014,kasperczyk_excitation_2015}. Notice that, when scanning the Si-NRs, the Raman spectrum due to two optical phonon processes can be detected (grey band in the spectrum), which is helpful defining precisely the position of the Si nanoantennas.

%% ------------------------ Figure ------------------------
\begin{figure*}[p]
\includegraphics[width=0.9\linewidth]{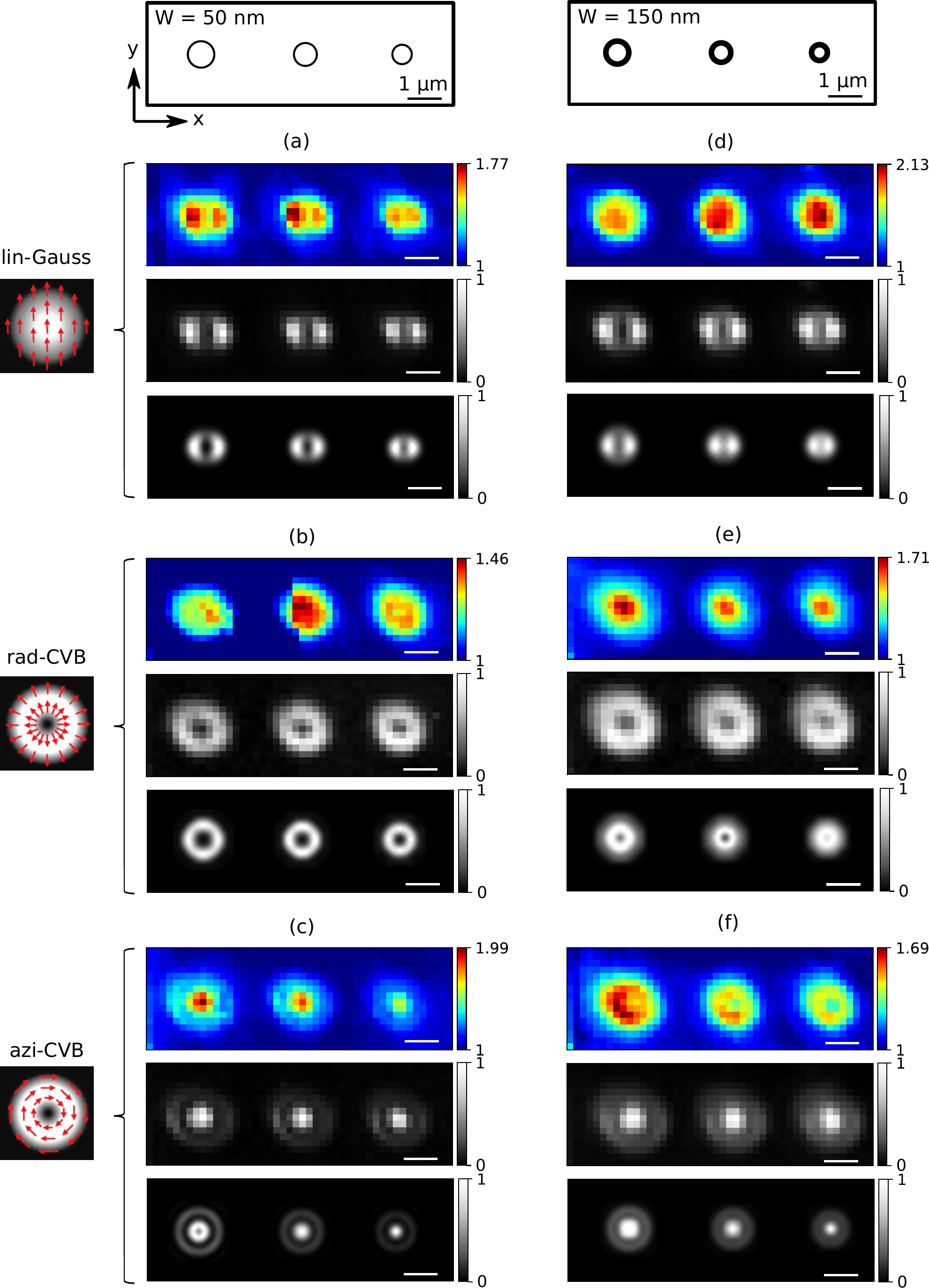}
\caption{\label{fig:fig2} \textbf{Photoluminescence and Raman mappings.} Left column: Si-NRs of  $W$~=~50 nm width (radius $R$ = 500, 450, and 400 nm from left to right) excited by, from top to bottom, (a) lin-Gauss, (b) rad-CVB and (c) azi-CVB. For each excitation are shown the PL mapping (top row), the Raman scattering mapping (middle row), and the calculated electric field intensity distribution inside the Si-NR. The PL maps are normalized to the signal far from the Si-NRs. Both experimental Raman and calculated electric field intensity maps are normalized to local maximum. Right column: same as left column for Si-NRs of $W$~=~150 nm width. The schemes showing the different nanorings is given on top of the mappings described above.}
\end{figure*}
%% --------------------------------------------------------

For electro-dynamical simulations, we used either the pyGDM toolkit implementing the Green Dyadic Method \cite{wiecha_pygdmpython_2018}, or the Meep software package based on the finite-difference time-domain (FDTD) method \cite{oskooi_meep_2010}.

\section*{Experimental results}

%There is no obvious difference using the 528 nm or 532 nm wavelength. 
In the following, we address the Si Raman signal and the total PL intensity given by the integrated intensity of the three contributions. Typical PL and Raman scattering (RS) mappings are presented in figure~\ref{fig:fig2}. The main results that can be underlined are: (i) the RS maps are dependent on the excitation polarization only, and (ii) the  PL maps not only depend of the excitation polarization but also on the Si-NR width $W$.\\

\indent \textbf{Raman mappings of Si nanorings.} 
As shown in figure~\ref{fig:fig2}, the Raman mappings give the same signature for all nanorings for a given laser polarization. There are two lobes for the (vertical) linear polarization, a doughnut shape for the radial polarization, and a centered spot for the azimuthal polarization. The fact that the signal intensity map depends only on the laser polarization for a nanoring shape leads us to the assumption that the Raman signal is driven by the local electric field inside the nanoring \cite{dmitriev_resonant_2016, raza_raman_2021}. To prove this, we calculated by FDTD the electric-field intensity inside the Si-NR in each point of the raster scan of the laser beam. The simulations are also displayed in figure~\ref{fig:fig2} below the experimental RS maps.\\
Both the experimental Raman maps and the calculated internal field intensity maps are in perfect agreement, validating our hypothesis.\\

%% ------------------------ Figure ------------------------
\begin{figure}[h!]
\includegraphics[width=0.4\textwidth]{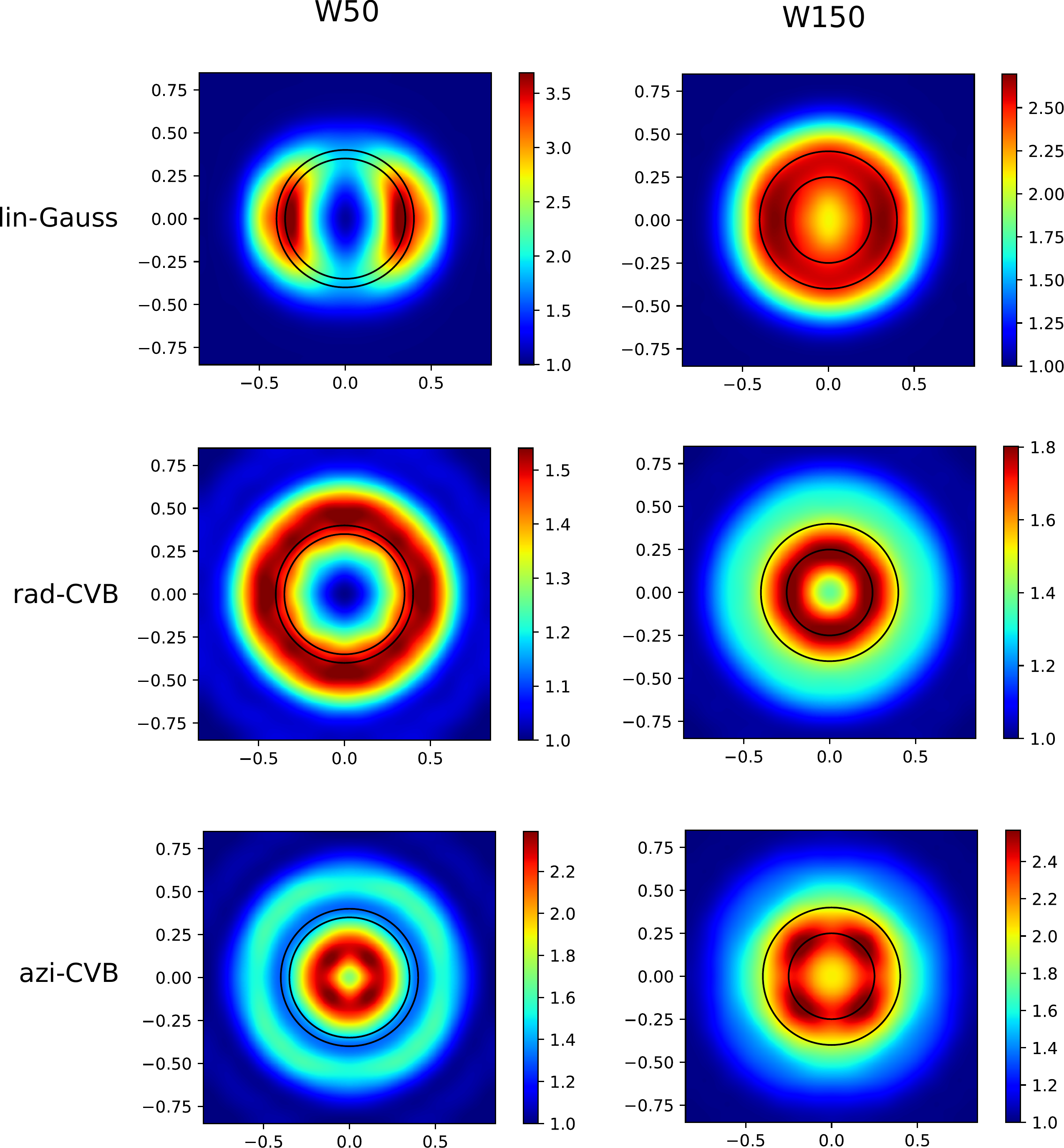}
\caption{\label{fig:fig3} \textbf{FDTD simulations of normalized near-field electric intensity distribution in the Eu$^{3+}$-doped layer.} From top to bottom, normalized near-field electric intensity distribution obtained by raster scanning the lin-Gauss, rad-CVB and azi-CVB excitation, respectively. The position in height of the raster scan is along the Eu$^{3+}$:Gd$_{2}$O$_{3}$ layer (10 nm above the silica substrate or the HSQ layer covering the 400 nm radius Si-NR).  Left columns: Si-NRs of $W$~=~50 nm. Right columns: Si-NRs of  $W$~=~150 nm  (The nanoring contour is plotted as black line).}
\end{figure}
%% -------

\indent \textbf{Photoluminescence mappings of Eu$^{3+}$:Gd$_{2}$O$_{3}$ deposited films on nanorings.} 
Confirming AFM images and previous experiments \cite{wiecha_enhancement_2019}, the cluster-deposited film is highly homogeneous as shown by the constant intensity observed in PL mappings far from the Si nanostructures. \\
\indent There is a slight enhancement of the Eu$^{3+}$ emission intensity above the Si-NR, but more important we point out a strong modification of the spatial distribution of the PL intensity as function of the excitation polarization. This spatial distribution is also strongly dependent on the nanoring dimensions in the case of the rad-CVB and azi-CVB excitations (See figure~\ref{fig:fig2}).

For linear polarized Gaussian beam, the PL enhancement evolves from two lobes for $W$~=~50 nm to a large central maximum with increasing width $W$.
For radial polarization, the PL enhancement evolves from a doughnut shape for $W$~=~50 nm to a narrow dotted maximum in a the ring center with increasing width $W$.
This behavior is fully reversed in the case of azimuthal polarization, with an evolution from narrow dotted maximum for $W$~=~50 nm to the doughnut shape with increasing width $W$.

To understand these experimental results, we calculated using FDTD the electric near-field intensity distribution by raster scanning the different CVBs inside the Eu$^{3+}$-doped film ($i.e.$ following a profile at 10 nm above the substrate surface and the Si-NR top surface including the HSQ layer). The near-field intensity distribution is calculated for each laser spot center position of the raster scan. \\
The results are given in figure~\ref{fig:fig3} in the case of 400 nm radius Si-NRs with two selected widths $W$~=~50 nm and $W$~=~150 nm. There is a qualitative agreement between the near-field simulations of figure and the PL experiments. Indeed, in the case of the azi-CVB (bottom line of figures \ref{fig:fig2} and \ref{fig:fig3}) , there is an obvious tendency for the near-field intensity maximum to form a doughnut above the Si-NR contour for the Si-NR with the largest width $W$=~150 nm, to a reduced spot more centered inside the Si-NR contour for the narrowest $W$=~50 nm. The behavior is reversed in the rad-CVB case (middle line of figures \ref{fig:fig2} and \ref{fig:fig3}): large doughnut shape for $W$=~50 nm and centered spot for $W$=~150 nm. In the lin-CVB case, we observe an evolution from the two lobes for the narrowest $W$ to a larger central spot for the largest $W$ as in the PL maps.

As we chose the laser power to stay below the saturation regime (where the emission is driven by the LDOS) \cite{girard_generalized_2005, majorel_quantum_2020},  and that the excitation wavelength at 532 nm corresponds to an electric dipole transition \cite{kasperczyk_excitation_2015}, the good agreement with the electric near-field simulations shows that the PL is governed by the electric near-field distribution around and above the Si nanostructures. Our results also prove that the PL response can be tuned by the combination of Si nanoantennas and excitation beam shape and polarization. 

Depending on these parameters, it is thus possible to have a narrow spotted PL maximum, which can be seen in both cases of azi-CVB and $W$~=~50 nm diameter Si-NR (Fig.~\ref{fig:fig2}c), and rad-CVB  and $W$~=~150 nm diameter Si-NR (Fig.~\ref{fig:fig2}e). By increasing the objective NA or the local electric field intensity (for instance by decreasing both the nanoring and central hole diameters), we could expect an enhanced PL in a reduced spot size, improving considerably the spatial resolution. We point out that this effect could be even higher using the magnetic field with the appropriate excitation wavelength (527.5~nm) resonant with the $^{7}$F$_{0}$~$\rightarrow$~$^{5}$D$_{1}$ magnetic dipole transition in Eu$^{3+}$ \cite{kasperczyk_excitation_2015}, as the magnetic field intensity is usually much larger than the electric field inside dielectric nanostructures \cite{bakker_magnetic_2015, yang_boosting_2017}. There is however an experimental challenge of filling a hole of about 50 nm in diameter by the emitters.\\

\begin{figure*}
\includegraphics[width=1.0\linewidth]{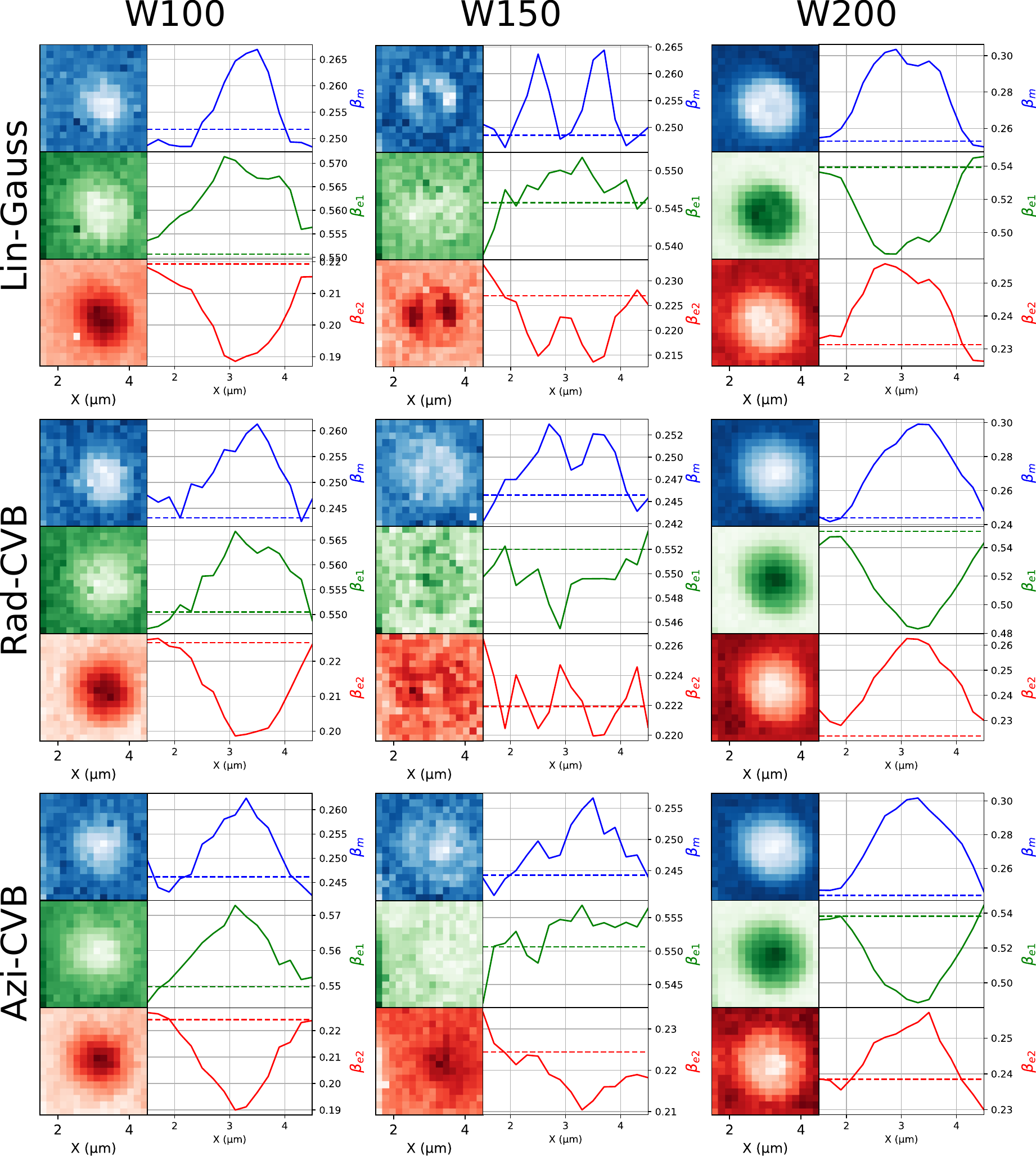}
\caption{\label{fig:fig4} \textbf{Branching ratios.} Branching ratio profiles for the different  transitions so called MD ($\beta_{m}$, blue), ED1 ($\beta_{e1}$, green) and ED2 ($\beta_{e2}$, red). The 1D profiles are taken along an horizontal line of the 2D mappings. The dashed lines in the spectra correspond to the BR of Eu$^{3+}$-doped film without any Si-NR antenna influence, referred to as BR$_{ref}$. The positions where all BRs coincide with their respective Br$_{ref}$ values correspond to positions between Si-NRs.}
\end{figure*}

\indent \textbf{Branching ratios and influence of radiative LDOS and light collection.}
After having investigated how the total PL intensity could be controlled by the excitation beam, we now focus on the influence of the system on the different competing emission channels, referred to as MD, ED1 and ED2 centered around 590 nm, 610 nm and 700 nm, respectively (See the experimental set-up section and figure~\ref{fig:fig1}).
The usual parameters used to describe how the emission channels are redistributed are the branching ratios, given by:
\begin{equation}\label{eq:branching ratio}
    \beta_{i} =  I_{i}/ I_{tot} = \Gamma^{rad}_{i}/\Gamma^{rad}_{tot}, 
\end{equation}
where $I_{i}$ is the integrated intensity of the transition labeled $i$ (MD, ED1, or ED2), $I_{tot}$ is the sum of the the three contributions, $\Gamma^{rad}_{i}$ is the radiative rate of the transition labeled $i$, and $\Gamma^{rad}_{tot}$ is the sum of the radiative rates of each transition.
This method allows to connect the intensity measured in stationary PL experiments to the radiative electric and magnetic LDOS \cite{sanz-paz_enhancing_2018, rabouw_europium-doped_2016, chacon_vectorial_2022}.\\
We show in figure~\ref{fig:fig4} the BR maps and 1D profiles for a Si-NR of 400 nm radius.
Though the influence of the excitation beam polarization is still visible, we point out the additional influence of the radiative LDOS as the BRs are affected as function of the Si-NR width. For the smallest $W$ (100 nm and below, left column of figure~\ref{fig:fig4}), the MD and ED1 emitting transitions tend to be favored to the detriment of ED2 transition. For the largest $W$~=~200 nm width, the MD is still enhanced while ED1 is now decreased to the benefit of ED2 (right column of figure~\ref{fig:fig4}). It seems that the $W$~=150 nm case corresponds to the transition between the extreme behaviors described above. Notice that the same behavior is found similar for different Si-NR radii (500-350 nm range) for a given $W$ value. The global behavior is also reproducible on different samples. The calculated magnetic and electric LDOS (not shown) do not fit the experiments, as the BRs depend only on the part of the radiative decay that enters the microscope objective \cite{rabouw_europium-doped_2016}. According to equation \ref{PL_intensity}, more complex simulations are needed taking into account the LDOS $\phi_{em}(\mathbf{r},\omega_{em})$, and the emission directivity and detection geometry $C_{coll}(\mathbf{r},\omega_{em})$.

We thus show that it is possible to shape the PL spatially, favoring different hot spots as function of the excitation beam polarization, while either enhancing or quenching a specific emission line of the Eu$^{3+}$-doped film.

%% ------------------------ Figure ------------------------
%% -------

%% ------------------------ Section ------------------------
\section*{Conclusions}

In conclusion, we investigated the photoluminescence mappings of Eu$^{3+}$-doped cluster deposited films on Si nanorings using different cylindrical vector beams. In addition to the photon LDOS, we show that the excitation beam shape and polarization have an important effect.
Radial and azimuthal polarized beams exciting ring shape nanoantennas allow to tune the near-field hot spots, hence the local PL enhancement.
Our results show that it is possible to spatially shape the photoluminescence of quantum emitters coupled to dielectric nanoantennas by both the excitation and emission channels. More complex dielectric nanoantennas could be designed to be resonant at an absorption transition excited by a chosen cylindrical vector beam, and at a specific emitting transition while quenching others. Rare earth ions supporting magnetic transitions at both absorption and emission are very good candidates to benefit from a very high local  magnetic field enhancement inside resonant dielectric nanoantennas at the excitation wavelength, combined to magnetic and electric LDOS engineering to control the different emission channels.

%% ------------------------ Section ------------------------
\subsection*{Acknowledgements} 
We acknowledge funding from Agence Nationale de la Recherche under project HiLight (ANR-19-CE24-0020-01), and support by the Toulouse computing facility HPC CALMIP (grants p12167 and p19042), and by the LAAS-CNRS micro and nanotechnologies platform, a member of the French RENATECH network. ICB is partner of the French Investissements d’Avenir program EUR-EIPHI (17-EURE-0002).

\subsection*{Author contributions}
Each coauthor contributed.

\subsection*{Competing interests}
The authors declare no competing interests.

\end{document}